\documentclass[apj]{emulateapj}
\usepackage[colorlinks = true, linkcolor = blue, urlcolor  = blue, citecolor = blue, anchorcolor = blue]{hyperref}
\usepackage{hyperref}
\usepackage{mathtools}
\usepackage{amsmath}
\usepackage{wasysym}
\usepackage{savesym}
\savesymbol{tablenum}
\usepackage{siunitx}
\restoresymbol{SIX}{tablenum}

\newcommand\fnurl[2]{\href{#2}{#1}\footnote{\url{#2}}}

\shorttitle{Active longitude and CME occurrences}
\shortauthors{N. Gyenge et al.}

\begin{document}

\title{Active Longitudes and Occurrence of Coronal Mass Ejections}

\author{N. Gyenge\altaffilmark{1,2*} , T. Singh\altaffilmark{3}, T. S. Kiss\altaffilmark{1,4}, A. K. Srivastava\altaffilmark{3}, R. Erd\'elyi\altaffilmark{1}}

\thanks{\altaffilmark{*}e-mail: n.g.gyenge@sheffield.ac.uk}

\affil{\altaffilmark{1}Solar Physics and Space Plasmas Research Centre (SP2RC), School of Mathematics and Statistics, University of Sheffield\\ Hounsfield Road, Hicks Building, Sheffield S3 7RH, UK\\
\altaffilmark{2}Debrecen Heliophysical Observatory (DHO), Konkoly Observatory, Research Centre for Astronomy and Earth Sciences\\ Hungarian Academy of Sciences, Debrecen, P.O.Box 30, H-4010, Hungary\\
\altaffilmark{3} Department of Physics, Indian Institute of Technology (Banaras Hindu University), Varanasi, India\\
\altaffilmark{4} Department of Physics, University of Debrecen, Egyetem t\'er 1, Debrecen, H-4010, Hungary\\}

\begin{abstract} 

 The spatial inhomogeneity of the distribution of coronal mass ejection (CME) occurrences in the solar atmosphere could provide a tool to estimate the longitudinal position of the most probable CME-capable active regions in the Sun. The anomaly in the longitudinal distribution of active regions themselves is often referred to as active longitude (AL). In order to reveal the connection between the AL and CME spatial occurrences, here we investigate the morphological properties of active regions. The first morphological property studied is the separateness parameter, which is able to characterise the probability of the occurrence of an energetic event, such as solar flare or CME. The second morphological property is the sunspot tilt angle. The tilt angle of sunspot groups allows us to estimate the helicity of active regions. The increased helicity leads to a more complex built-up of the magnetic structure and also can cause CME eruption. We found that the most complex active regions appear near to the AL and the AL itself is associated with the most tilted active regions. Therefore, the number of CME occurrences is higher within the AL. The origin of the fast CMEs is also found to be associated with this region. We concluded that the source of the most probably CME-capable active regions is at the AL.  By applying this method we can potentially forecast a flare and/or CME source several Carrington rotations in advance. This finding also provides new information for solar dynamo modelling.

\end{abstract}

\section{Solar Non-Axisymmetric Activity}

Since the beginning of the last century the non-homogeneous spatial properties of solar activity have been studied extensively \citep{Carrington1863, Chidambara, Maunder1905, Losh, Bumba1965, Bumba69b}. These early investigations conjectured initially that the longitudinal distribution of sunspot groups or sunspot numbers shows non-homogeneous behaviour. These analyses concluded that there are preferred longitudes, where solar activity concentrates. 

Later, different approximations and assumptions were applied to understand the essence of this phenomenon. The topic soon became controversial \citep[see e.g.][]{Pelt06, Henney05}. Overall, three approaches can be distinguished. The first approach is the quasi-rigid structure model by \cite{Warwick66}. This model describes a constantly rotating frame which carries the persistent domains of activity \citep{Ivanov07}. \cite{Bogart82} applied an autocorrelation statistical method based on long-term sunspot number data. \cite{Balthasar83, Balthasar84} applied period-analysis to the Greenwich Photoheliographics Results (GPR). These studies concluded that the angular velocity of the quasi-rigid rotating frame varies. The angular velocity depends on the solar cycle, but during one cycle the angular velocity is constant.

The second approach, promoted by e.g. \cite{Becker55, Castenmiller86} and \cite{Brouwer1990} discovered the 'active nest' and defined it as a small and isolated area on the solar surface. Here, the enhanced longitudinal activities are considered as individual entities. These isolated entities can be absent for several rotations.

The third group of models assumes a migrating activity in the Carrington coordinate system. \cite{Berdyugina03, Berdyugina04, Berdyugina05} found persistent ALs under the influence of the differential rotation. \cite{Usoskin05} and \cite{Berdyugina06} introduced a 'dynamic reference frame'. This frame describes the longitudinal migration of active longitude in Carrington coordinate system and the frame has a similar dynamics to differential rotation. \cite{Usoskin07} concluded that the migration of the enhanced activity is just apparent. In the 'dynamic reference frame', the rotation of the active longitude remains constant and the active longitude itself is a persistent quasi-rigidly rotating phenomenon. \cite{Usoskin07} proposed that a seemingly migrating AL may occur as a result of interaction between the equatorward propagating dynamo wave and a quasi-rigidly rotating non-axisymmetric active zone. The role of differential rotation is also controversy. Various studies concluded that the differential rotation is not the reason of the migration of the AL \citep{Balthasar07, Juckett06, Juckett07, Gyenge14}.

In our previous work (i.e. \cite{Gyenge16}, hereafter GY16), we found evidence supporting the third group of models. Moreover, the migration of ALs does not appear to correspond very well to the form of the 11-year cycle as suggested by previous studies \citep{Usoskin05, Berdyugina06}. The half-width of the active longitudinal belt is fairly narrow during moderate activity but is wider at maximum activity. We also found that the AL is not always identifiable.

Several studies \citep{Warwick65, Bai03a, Bai03b} suggested that the spatial distributions of eruptive solar phenomena also show non-axisymmetric properties. \cite{Bai87, Bai88} analysed the coordinates of energetic solar flares based of 5 years of time period and concluded that  longitudinal spatial distribution is non-homogeneous. \cite{Zhang07b, Zhang08, Zhang15} concluded that the dominant and co-dominant AL contains 80\% of C- and X-flares. In GY16, we conducted a similar study based on four solar cycles and we did not find significant co-dominant activity; instead, we found that only the dominant AL contains 60\% of the solar flares.

The flares and CMEs could occur independently of each other. Numerous CMEs have associated flares but several non-flaring filament lift-offs also lead to CME \citep{Gosling76, Harrison95}. Furthermore, in the case of the 'stealth' CMEs there are no easily identifiable signatures to locate the source of the eruption on the solar surface \citep{Howard13}. Hence, separate flare - CME spatial distribution investigations are justified.

\section{CME and sunspot data}

We used the SOHO/LASCO HALO CME catalog by the \fnurl{CDAW data centre}{http://cdaw.gsfc.nasa.gov/CME_list/halo/halo.html}. The CME catalog spans over 20 years, i.e. between 1996 and 2016. This is the most extensive catalog that contains the source location of CMEs. Only halo CMEs are reported, i.e. their angular width is 360 degrees in C3 coronagraph Field of View. \cite{Gopalswamy09} describes the catalog in great details. The CME source is defined as the centre of the associated active region. The source of the CME is identified using SOHO EIT running difference images. Later, the ability of the STEREO mission to observe the backside of Sun was used to identify the source of back-sided halo CMEs \citep{Gopalswamy15}. This catalog also provides the space speed of CMEs which is the actual speed with which the CME propagates in the interplanetary space. The Plane of Sky speed obtained from the single SOHO viewpoint, converted into space speed using the cone model \citep{Xie04}.

The \fnurl{Debrecen Photoheliographic Data}{http://fenyi.solarobs.unideb.hu/DPD/ } (DPD) sunspot catalogue is used for estimating the longitudinal position of the AL. The catalog provides information about the date of observation, position and area for each sunspot. The precision of the position is 0.1 heliographic degrees and the estimated accuracy of the area measurement is $\sim$10 percent.

\section{Longitudinal sunspot distribution}

\begin{figure*}
	\centering
	\includegraphics[width=145mm]{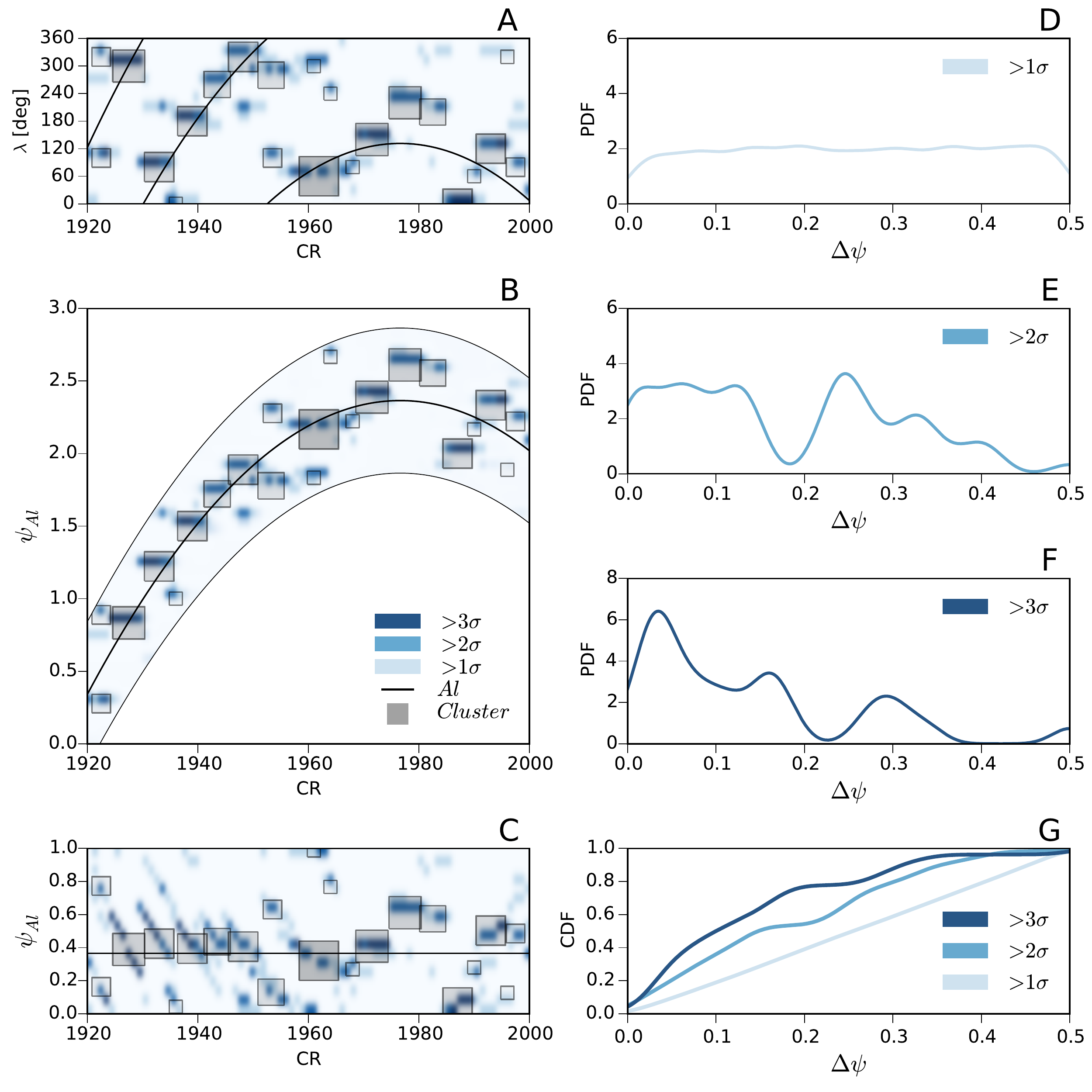}
	\caption{The Panels $A$, $B$ and $C$ show an example of the migration of the active longitude between $1920$ CR and $2000$ CR (01/03/1997-20/02/2003) based on data of the solar northern hemisphere. The shades of blue indicates the significance of sunspot group activity. The grey squares show high-density areas, i.e. the detected clusters. The solid black line represents the migration patch, fitted by the least-square-method and considering only the most significant (above 3$\sigma$) clusters. Panel $A$ shows the observed longitudinal distribution of sunspots in  Carrington coordinate system. Panel $B$, similar to Panel $A$ , depicts solar circumference (Carrington phase) repeated three times. Panel $C$ is the phase-corrected migration path. The parabolic migration pattern is now transformed to a constant line, and provides an insight into the non-homogeneous spatial property. The panels $A$ and $C$ use the same colour scale as defined in Panel B. The colour scale is displayed in the lower right corner of the Panel $B$. Panels $D$, $E$ and $F$ show the sunspot distribution around the active longitude (corresponding to $\Delta\psi=0$) with different significance levels taken for the entire time period and for both hemispheres. The horizontal axes ($\Delta\psi$ - Carrington Phase Difference) represent the shortest distance between the migration of enhanced longitudinal activity and given sunspot groups. Panel $G$ is the cumulative distribution of the above three PDF.}
	\label{al}
\end{figure*}

The first step of our identification procedure is to divide the solar surface by 18 equally sliced longitudinal belts. Hence, one bin equals to a zone with $20^{\circ}$ width. We take into account all sunspot groups from the moment when they reach their maximum area. This filtering criteria is chosen for the following reasons: Firstly, if we select all sunspot groups at every moment of time then the statistics will be biased by the long-lived sunspot groups. Secondly, the maximum area of the sunspot groups is a well-defined and easily identifiable moment. This is different from the used practice of considering only the first appearance of each sunspot group. Let us define the matrix $W$ by:

\begin{equation}
      W_{\lambda, \text{CR}} = \frac{A_{i, \text{CR}}}{ \sum_{i=1}^{n} A_{i,\text{CR}} }.
\label{W}
\end{equation}

Here, the area of all sunspot groups ($A_{i, \text{CR}}$) is summed up in each longitudinal bin for each CR. $ A_{i, \text{CR}}$ is divided by the summarised area over the entire solar surface ($ \sum_{i=1}^{n} A_{i,\text{CR}} $). The range of $W$ must be always between $0$ and $1$, depending on the local appearance of the activity. In the case of $W_{\lambda, \text{CR}} = 1 $, all of the flux emergence takes place in one single longitudinal strip. A $3 \sigma$ significance level threshold is applied to filter the noise. Moving average is also applied for data smoothing with a time-window of 3 CRs.

We standardised the matrix $ W_{\lambda, \text{CR}} $ (defined in Eq \ref{W}) by removing the mean of the data and scaling to unit variance. Then, cluster analysis was performed for grouping the obtained significant peaks. Here, the DBSCAN clustering algorithm was chosen which is a density-based algorithm. The method groups together points that are relatively closely packed together in a high-density region and it marks outlier points that stand alone in low-density regions (for details see \cite{Ester96}). The parameter epsilon ($=0.2$) defines the maximum distance between two points to be considered to be in the same group. The parameter m ($= 3$) specifies the desired minimum cluster size. Clusters, containing less than three points, were omitted. The longitudinal location of the clusters ($\lambda_{cluster, \text{CR}}$) represent the position of the AL.

Panel $A$ of Figure \ref{al} shows an example of the initial identification steps outlined above. The sample time period covers 6 years, and it  corresponds to 80 CR between $1920$ CR and $2000$ CR. The quantity $W$ is represented by the shades with blue colour. The dark blue regions denote the significance ($3 \sigma$)  presence of activity. The brighter shades stand for a weaker manifestation of activity ($2 \sigma$ and $1 \sigma$). The grey squares shows the sunspot group clusters. 

In Panel $B$ of Figure \ref{al}, the Carrington longitudes of the most significant cluster is transformed into Carrington Phase for each CR:

\begin{equation}
	\psi_{\text{Al, \text{CR}}} = \frac{\lambda_{\text{cluster, \text{CR}}}}{360} .
	\label{Crp}
\end{equation}

The range of the quantity $\psi_{\text{AL}}$ must be between $0$ and $1$ where $\psi_{\text{AL}}=1$ represents the entire circumference of the Sun. In Panel $B$ of Figure \ref{al}, Panel $A$ is repeated three times so that we are able to track the migration of the activity thought the phases. To the data Panels $A$ and $B$, we applied a polynomial least squares fitting based on multiple models. Linear, quadratic, cubic and higher-order polynomial models were tested. The quadratic or parabolic regression ($ A x^2 + Bx - C$) shows the best goodness of fit, hence this model is chosen. Table \ref{Coefficient} shows the coefficients and uncertainties.

The shape of migration clearly follows parabolic-shaped path as found in several earlier studies \citep{Berdyugina06, Berdyugina03, Zhang11a, Usoskin07, Usoskin05, Gyenge14}. 

\begin{table*}
	\centering
		\caption{The coefficients and uncertainties of the parabolic functions.}
		\label{Coefficient}
		\begin{tabular}{lllll}
			Solar Cycle & Hemisphere  & Coefficient A & Coefficient B & Coefficient C \\
			23 & North & $  \num{-4.32e-4} \pm \num{5.44e-5}$ & $ \num{1.71e0} \pm \num{2.15e-1}$  &  $ \num{1.69e3} \pm \num{2.13e2}$ \\
			24 & North & $ \num{-5.61e-4} \pm \num{1.78e-4}$ & $ \num{2.37e0} \pm \num{7.47e-1}$ & $ \num{2.51e3} \pm \num{7.83e2}$\\
			23 & South & $ \num{-5.62e-4} \pm \num{6.62e-5}$ & $ \num{2.23e0} \pm \num{2.62e-1}$ &$ \num{2.20e3} \pm \num{2.59e2}$\\
			24 & South & $ \num{-5.16e-4} \pm \num{1.85e-4}$  & $ \num{2.18e0} \pm \num{7.77e-1}$ & $ \num{2.31e3} \pm \num{8.12e2}$                 
		\end{tabular}
\end{table*}

\begin{figure*}
	\begin{center}
		$\begin{array}{clcr}
		\includegraphics[width=77mm]{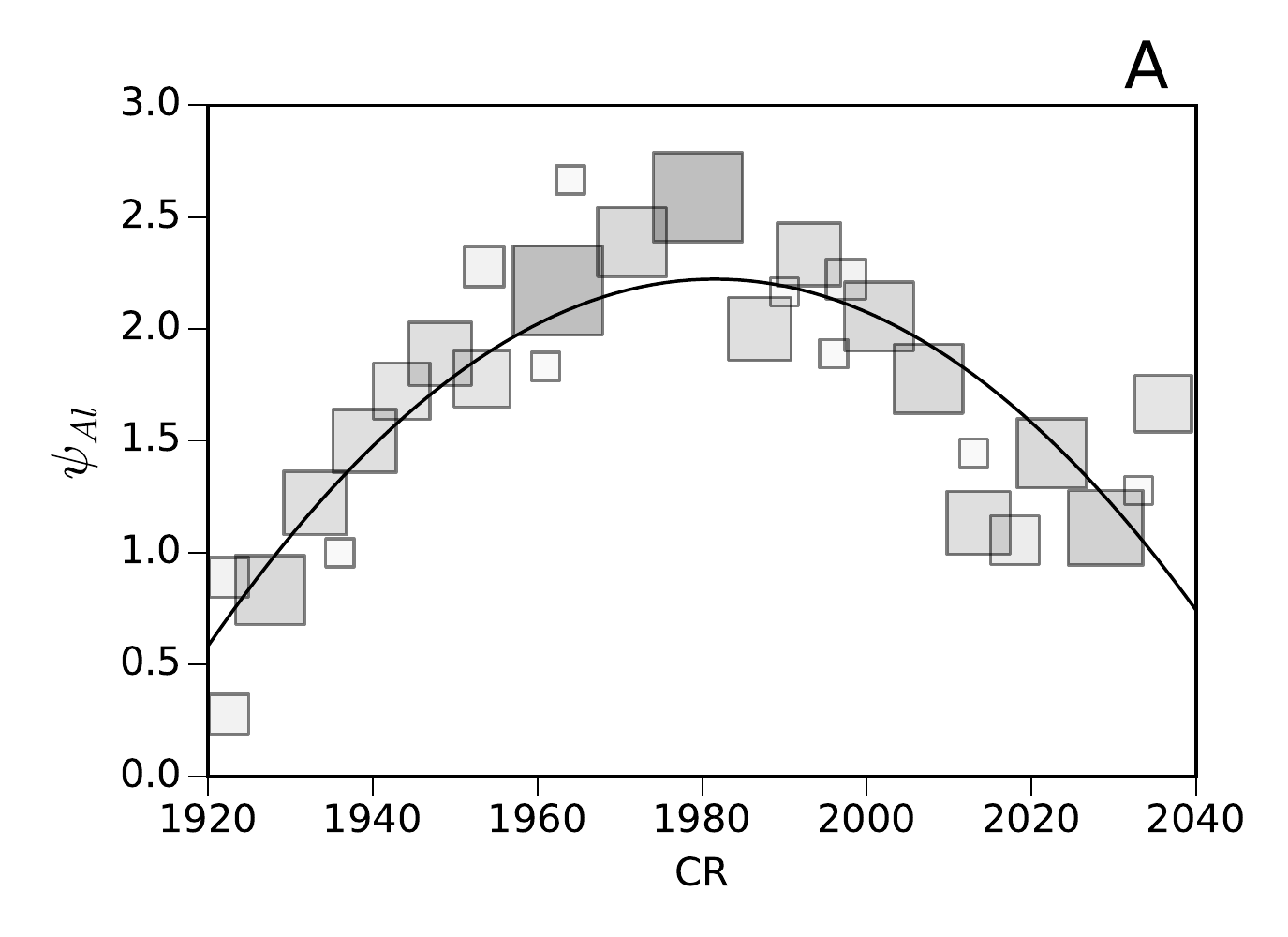} &
		\includegraphics[width=77mm]{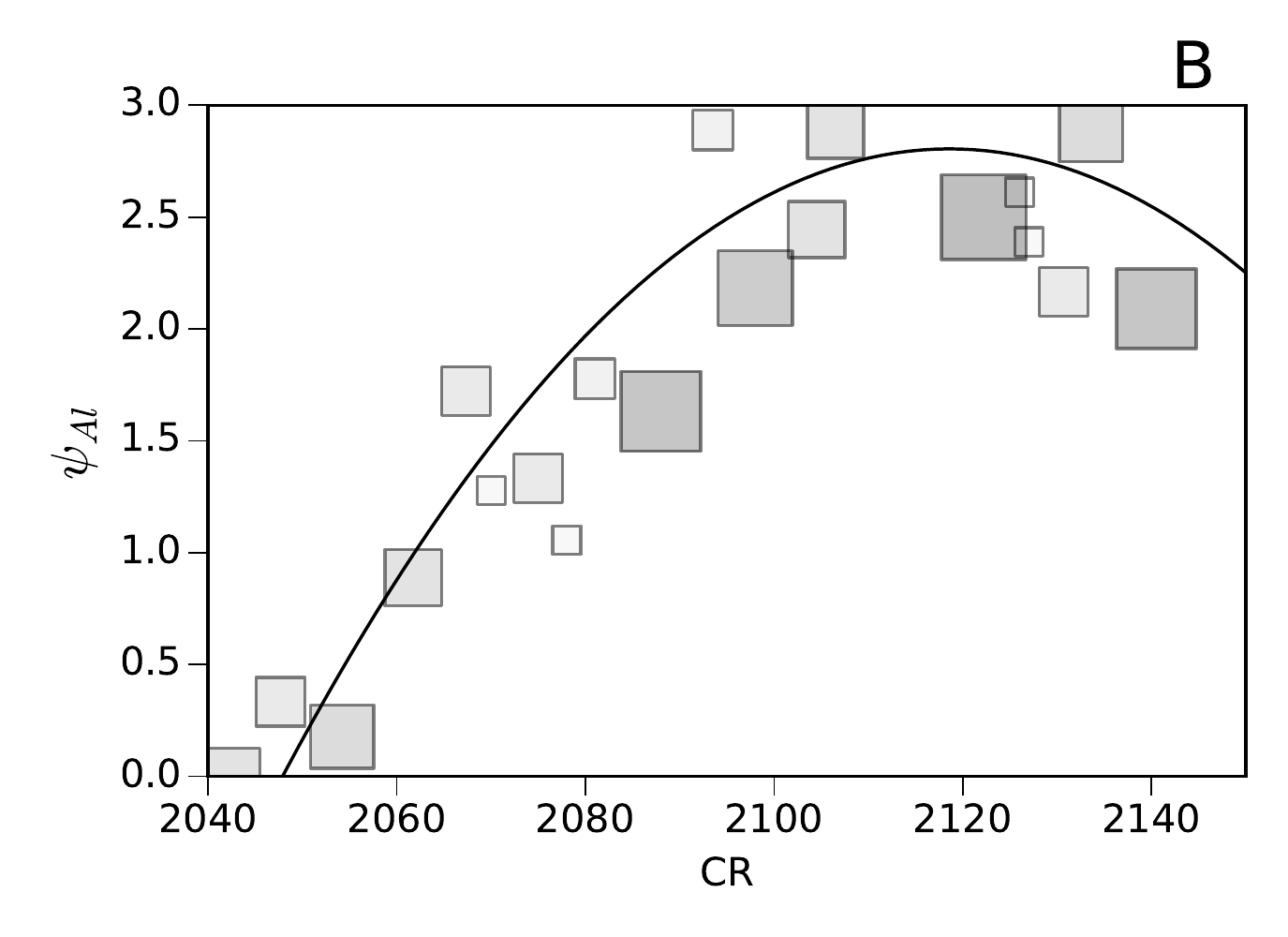} \\
		\includegraphics[width=77mm]{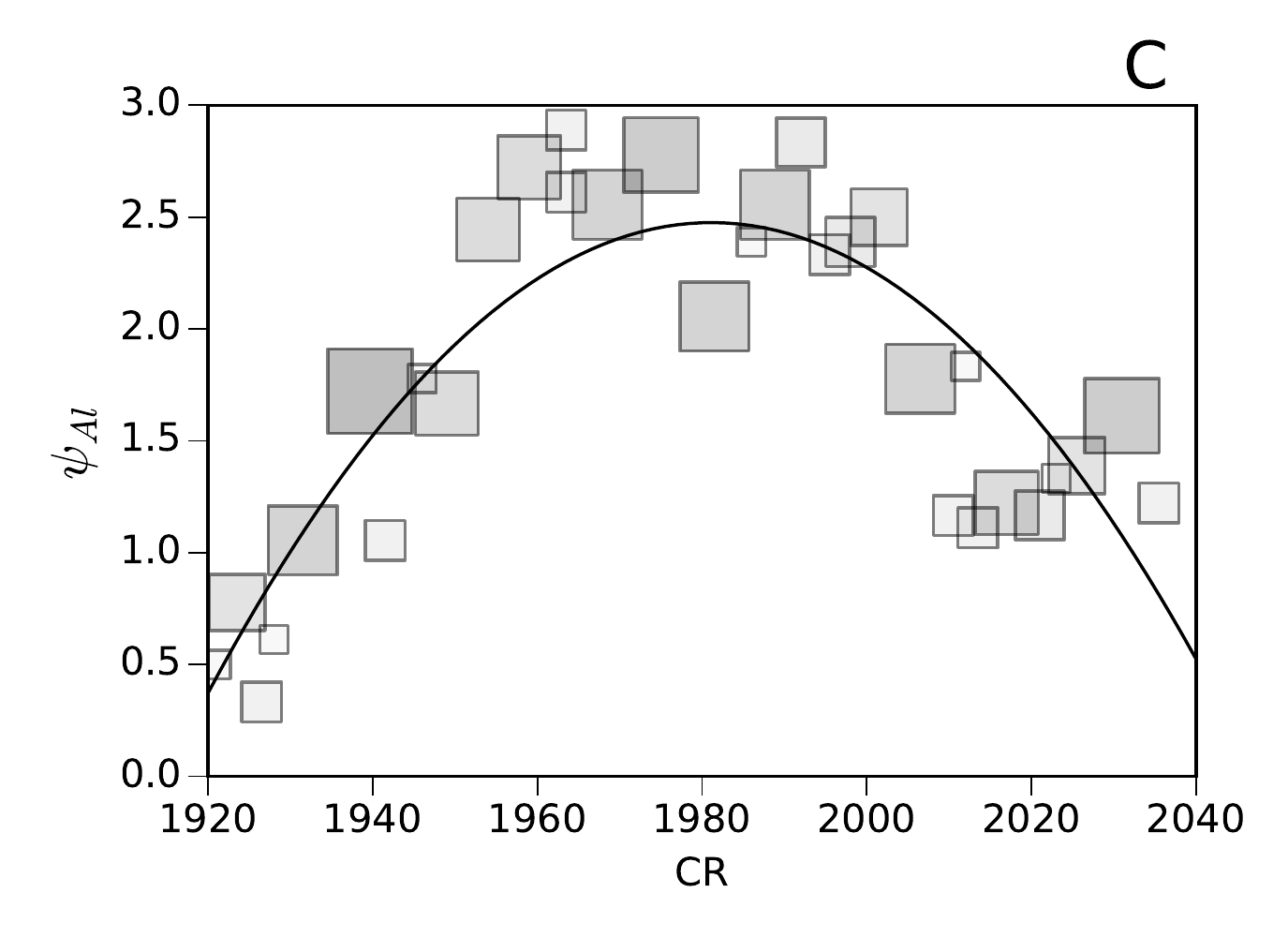} &
		\includegraphics[width=77mm]{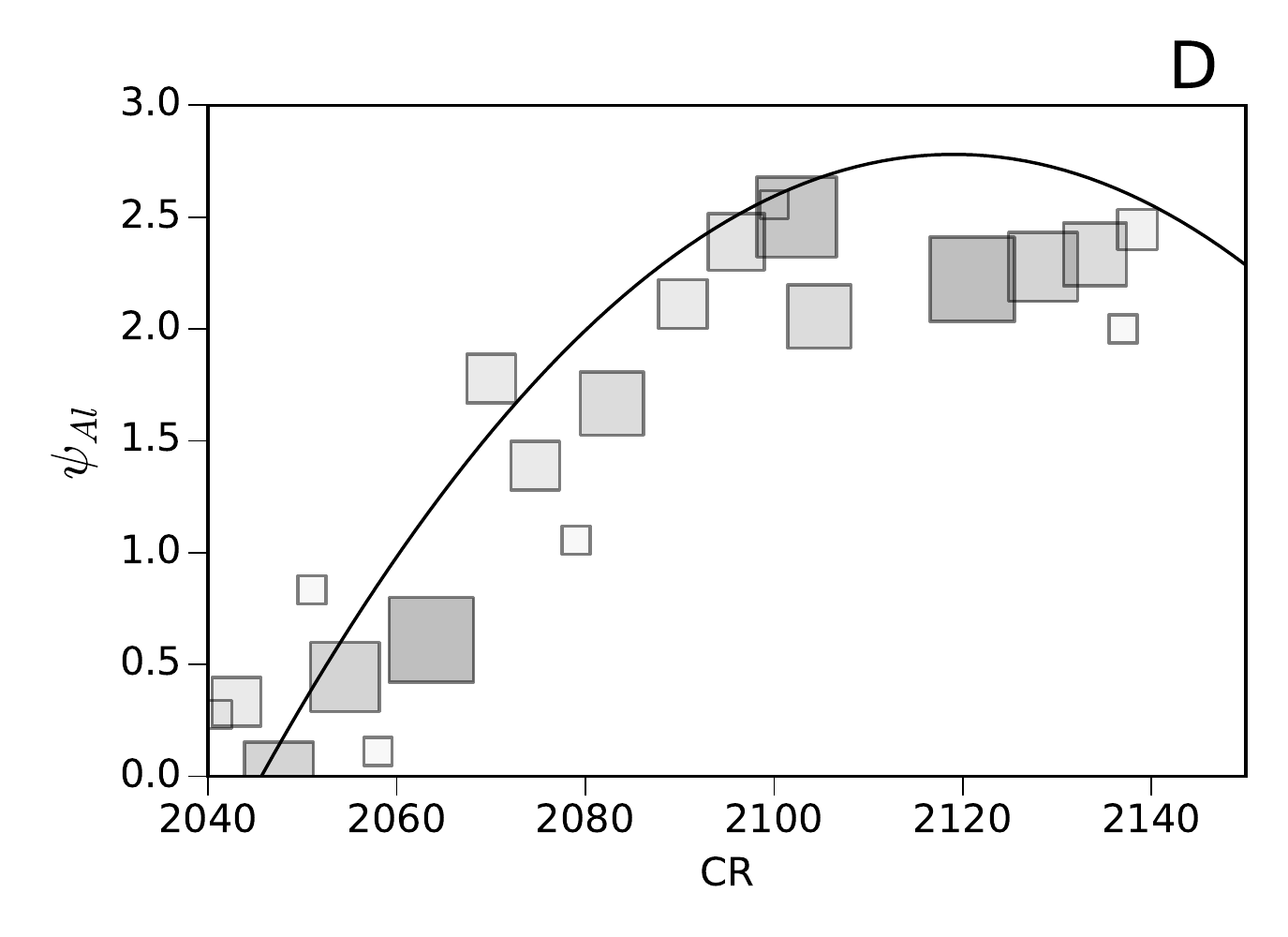}
		\end{array}$
		\end{center}
		\caption{The four panels demonstrate the phase-corrected migration path based on the entire analysed period. The Panels $A$ and $C$ show the AL in Solar Cycle 23 and the Panels $B$ and $D$ in Solar Cycle 24. Panels $A$ and $B$ demonstrate the northern hemisphere data and Panels $C$ and $D$ show the southern hemisphere. The grey squares visualise the result of the DBSCAN algorithm, i.e. the detected and significant sunspot group clusters. The quadratic model function is applied. The coefficients and uncertainties are given in Table \ref{Coefficient}. }
	\label{AL_ent}
\end{figure*}
  
In Panel $C$ of Figure \ref{al}, let us now plot the parabola-shaped migration path in a coordinate system that follows the parabolic shape of Panel $B$. Hence, the active longitude is easily recognisable without repeating Carrington Phases. Several new features are noticeable in Panel $C$; parallel aligned and tilted lanes between CR 1930 and CR 1950. These shapes are artificial, created by the coordinate system transformation. In the new dynamic coordinate system the AL stands still as represented by the black regression line.

Panels $D, E$ and $F$ of Figure \ref{al} show the kernel density estimation (KDE) of the spatial difference between the AL ($\psi_{\text{AL,CR}}$) and the longitudinal position of individual sunspot groups ($\psi_{\text{CR}}$) in Carrington phase difference ($\Delta\psi$, see Eq. \ref{Crd}) applied to the entire period investigated (i.e between $06/01/1997$ and $30/12/2015$) and for both hemispheres.

\begin{align}
	\begin{split}
		&\Delta\psi^{*} =  \left|\psi_{\text{Al, \text{CR}}}-\psi_{\text{\text{CR}}} \right|,\\
    	\Delta\psi = & 
		\begin{cases}
    		 \Delta\psi^{*} ,           & \text{if }  \Delta\psi^{*} \leq  0.5,\\
    		1 - \Delta\psi^{*},        & \text{if }  \Delta\psi^{*}  >    0.5.
		\end{cases}
	\label{Crd}
	\end{split}
\end{align}

The KDE is a non-parametric method to estimate the probability density function \citep{Connolly00}. We used a Gaussian kernel function. The optimal value of the Gaussian kernel bandwidth is $0.02$. If $\Delta\psi=0$ the sunspot group is located the AL. In case of $\Delta\psi=0.5$, the location of the sunspot groups are shifted by 180 degrees from the  position of the AL. This shifted position is marking the area of the co-dominant AL. The three KDEs are obtained from  different sunspot group datasets with different threshold levels. Panel $F$ of Figure \ref{al} shows the most significant regions ($3 \sigma$ significance threshold was applied) which tend to be formed near the active longitude. The less significant ($2 \sigma$) sunspot groups show a more disperse distribution. Below the $1 \sigma$ threshold, the KDE does not show obvious peaks. Panel $G$ of Figure \ref{al} depicts cumulative distribution functions (CDF) of the three KDEs. Note that $70\%$ of the most significant sunspot groups appear closer than $\Delta\psi=0.17$. This value corresponds to a longitudinal zone with $\pm 60$ degrees of width around AL. The Figure \ref{AL_ent} show the migration path of the AL based on the entire analysed period. The coefficients and uncertainties of the employed quadratic model functions are displayed in Table \ref{Coefficient}. 

\section{Complexity properties of AL sunspots}

\subsection{Separateness parameter}

\begin{figure*}
	\center
	\includegraphics[width=142mm]{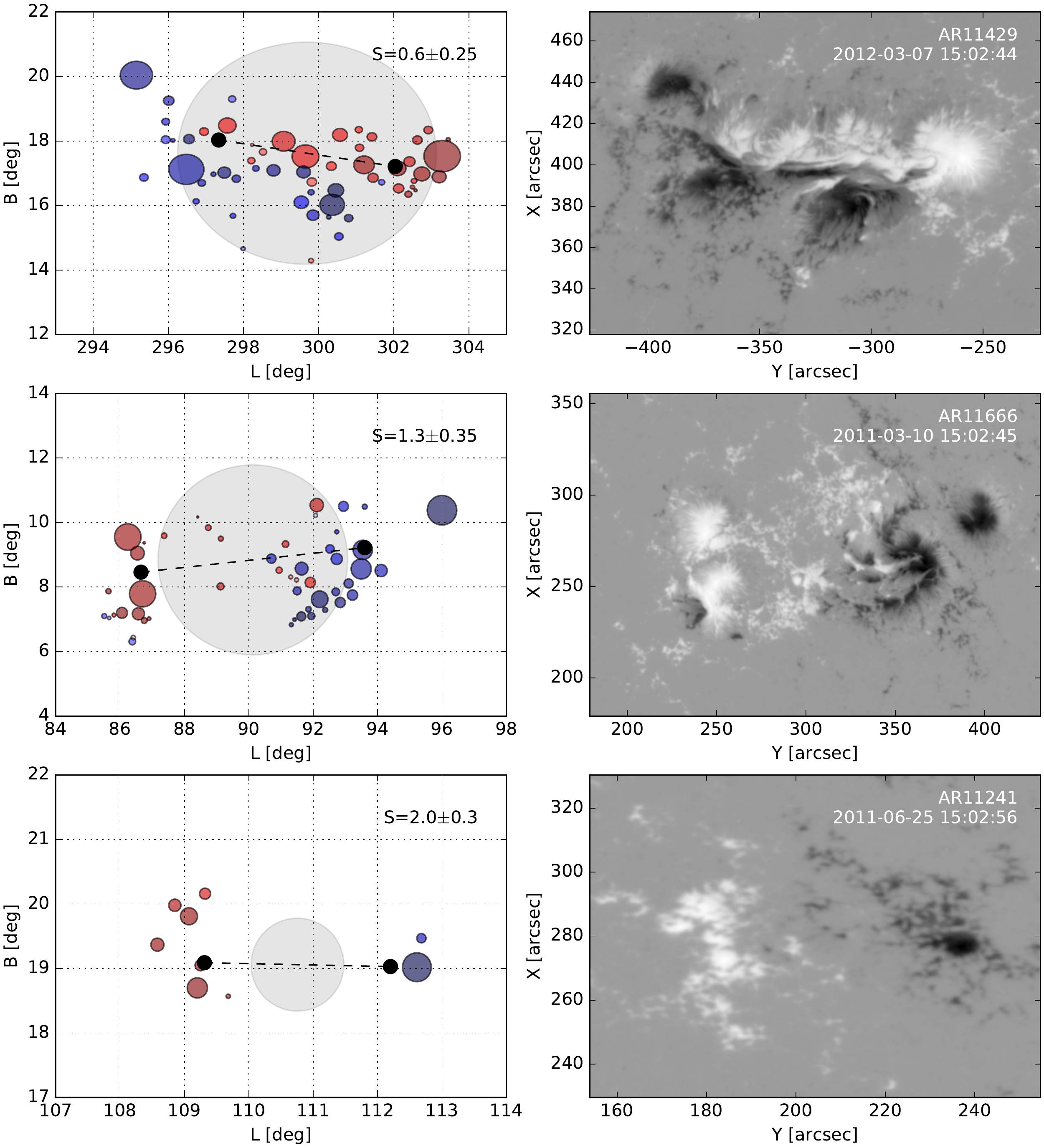}
	\caption{Representation of the reconstructed sunspot groups (left-hand side) using the Debrecen Sunspot Data (DPD) catalogue applied to active regions NOAA 11429 (07/03/2012 15:02:44),  NOAA 11666 (0/03/2011 15:02:45),  NOAA 11241 (25/06/2011 15:02:26). The polarities of the spots are distinguished by the different colours (red and blue). The grey colour is the hypothetical circle, having summarized area of the group. The dashed line is the distance between the following and leading subgroups. The area is measured in MSH (millionths of solar hemisphere). The sunspots are corrected for foreshortening. On the right-hand side, the magnetograms of the example sunspot groups are displayed.}
	\label{Sep}
	\center
\end{figure*}

The relationship between flare occurrences and the morphological properties of sunspot groups is widely accepted and is reported in numerous studies \citep[e.g.][]{Schrijver07, Cui06, Mason10, korsos14, korsos15b}. The first classification scheme \citep{Hale19} is published by \cite{Waldmeier38}. The scheme examines the role of the size and morphology of sunspot groups in relation of determining the capacity of their flare-productivity. The system was modified by \cite{Waldmeier47} and is known today as the modified Z\"urich classification system \citep{Kiepenheuer53}. The classification was further developed by \cite{McIntosh90} and its version is still in use widely today. Later the classification was automated by \citep{Colak08}. However, the classification of the  Z\"urich or McIntosh system is still subjective; further there are just a limited number of classes \citep{Bornmann94}. 

The separateness parameter ($S$) is used to reveal the morphological properties of the sunspot groups near and far from the AL. This parameter was introduced by \cite{korsos16} and its investigation and application to flares showed that the separateness parameter can be a numerical indicator/precursor besides the traditional (Z\"urich, McIntosh and Mount Wilson) classifications of sunspot groups. The parameter $S$ is considered as an indicator for the potential flaring outbreak and CME capability of active regions.

The separateness parameter is determined by the angular distance between the area-weighted centres of the leading and following subgroups divided by the angular diameter of a hypothetical circle whose area is equal to the total area of all umbrae constituting the sunspot group.

The angular distance is the shortest distance between two points on the surface of a sphere. The distance (in degrees) between the leading and following subgroups is provided by the spherical law of cosines:

\begin{align}
	\begin{split}
		\Delta \theta ={}& 2 \arcsin{ \bigg[ \sin^2( \frac{ \left| B_{l}-B_{f}\right|} {2} ) + } \\
		& + \cos(B_{l})  \cos(B_{f})   \sin^2( \frac{ \left| L_{l}-L_{f}\right|} {2} ) \bigg] ^{\frac{1}{2} } .
	\end{split}
\label{eq1}
\end{align}

Here, $B$ and $L$ refer to the heliographical latitude and longitude of $l$ leading and $f$ following subgroups. If the absolute difference is greater than 180 ($\left| L_{l}-L_{f}\right| > 180$), then the absolute difference is $360 - \left| L_{l}-L_{f}\right|$.

The corrected area of individual sunspots  ($A^{*}$) in millions of solar hemisphere is converted to $Mm^2$, using:

\begin{equation}
	A=\frac{1}{2} (4\pi R_{\text{Sun}}^2)10^{-7} A^{*},
\label{eq2}
\end{equation}

where $R_{\text{Sun}}$ is in $Mm$. The total sunspot group area ($T$) is calculated. The number of sunspots in a certain sunspot group is represented by the quantity $n$. The total area ($Mm^2$) means the summed up area of the individual sunspots:

\begin{equation}
	T=\sum_{i=1}^{n} A_{i}.
\label{eq3}
\end{equation}

The diameter of an individual sunspot group, ($\Delta \Omega$), is estimated by:

\begin{equation}
	\Delta \Omega=\frac{2\sqrt{T / \pi}}{2(R_{\text{Sun}} \cos(\frac{1}{2}(B_{l}+B_{f}))\pi}360^o.
	\label{eq4}
\end{equation}

The numerator is the diameter ($Mm$) of a hypothetical circle whose area is equal to the $T$ total area. The denominator represents the circumference of small circle ($Mm$), which connects all locations with a given latitude. The fraction is multiplied by $360$ degrees, equals to the angular distance between the endpoints of the sunspot group diameter in degrees.

Finally, let us define the dimensionless separateness parameter:

\begin{equation}
	S=\frac{\Delta \theta}{\Delta \Omega}.
	\label{eq5}
\end{equation}

In Figure \ref{Sep}, typical active regions NOAA $11429$, $11666$ and $11241$ (from top to bottom) are selected to demonstrate the usefulness of the parameter $S$. The visualisation of the active regions is plotted on the left-hand side. The blue and red colours distinguish the different magnetic polarities, the radius of a circle represents the area of the spot. The black dots indicate the weighted average position of the leading and following subgroups. The black dashed line between the black dots is the calculated angular distance ($\Delta \theta$), described by the Eq. \ref{eq1}. The grey circle around the spots is the hypothetical circle whose area is equal to the total area of all umbrae constituting the sunspot group ($\Delta \Omega$), defined by Eq. \ref{eq4}. Panels on the right-hand side are the snapshots (HMI magnetogramm by SDO) of the active regions.

The upper two panels of the Figure \ref{Sep} are visualising a complex sunspot group (namely, NOAA 11429). This sunspot group is a beta-gamma-delta magnetic configuration according to the Mount Wilson classification. The AR is extremely complex, having umbrae of opposite polarity within the same penumbra. The calculated separateness parameter is $0.6 \pm 0.25$ (Eq. \ref{eq5}). The middle panels of Figure \ref{Sep} show NOAA Active Region 11666, which is a moderate complex sunspot group ($S = 1.3 \pm 0.35$). The bottom panels display NOAA Active Region 11241, a less complex bipolar sunspot group ($S = 2.0 \pm 0.30$)

Based on the study by \cite{korsos16}, there is a hight risk of X-class flare and/or fast CME occurrence(s) if $S < 1$. There is a moderate risk of flaring (M-class) and/or CME occurrence(s) if $S > 1$ and $S < 2$. In case of $S>3$, only bipolar sunspot groups appear with relatively simple morphological properties. These ARs have a rather low probability of a significant flaring (above GOES C-class) or CME activities.

\subsection{Separateness parameter within AL}

\begin{figure*}
	\centering
	\includegraphics[width=150mm]{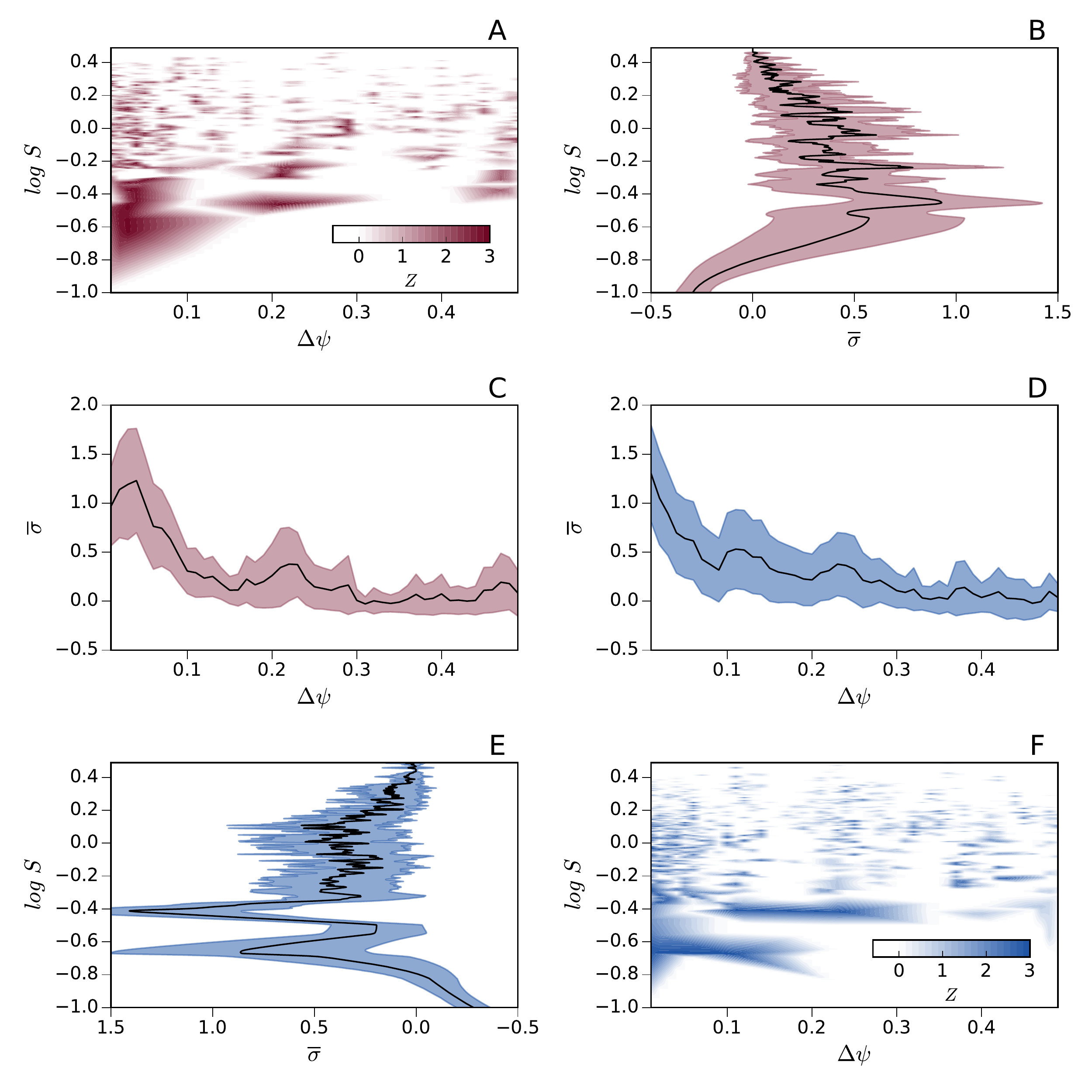}
	\caption{The separateness - phase - standard score statistic based on the northern and souther hemisphere, indicated by blue and red colours. The panels $A$ and $E$ shows the separateness of sunspot groups versus the longitudinal location of the active region. The shade of red and blue colours indicate the positive standard score. The Panels $B$, $C$, $D$ and $E$ are the cross-section PDF, produced by the KDE method.}
	\label{cmp_stat}
\end{figure*}

In this section, we investigate the longitudinal spatial distribution of the parameter $S$ (Eq. \ref{eq5}) and the area of the investigated sunspot groups. The spatial distance of sunspot groups ($\Delta\psi$) is defined by Eq. \ref{Crd}.

The raw area measures are converted to standard scores (standardized statistics). The standard score is a dimensionless quantity calculated by subtracting the average of sunspot group area ($\overline{T}$) from a given group area $T_{i}$ and dividing by the sample-corrected standard deviation ($\sigma(T)$) of the area data:

\begin{equation}
	Z_{i} = \frac {T_{i} - \overline{T}}  {\sigma(T)}.
	\label{z_score}
\end{equation}

Spatial multi-variable (linear) interpolation is applied to $f(S, \Delta\psi, Z)$. The method results in a regular matrix ($M$) form unstructured 3-dimensional data. The range of $\log S$ is $[-1, 0.5]$, and is divided by $1500$ equal bins ($n$). The range of $\Delta\psi$ $[0, 0.5]$ is divided by 500 bins ($m$), i.e.

\begin{equation}
	M=
		\begin{bmatrix}
			Z_{1,1} &  Z_{1,2} & ... & , Z_{1,n} \\
			Z_{2,1} &  Z_{2,2} & ... & , Z_{2,n} \\
			... &  ... & ... & , ... \\
			Z_{m,1} &  Z_{m,2} & ... & , Z_{m,n} 
		\end{bmatrix}.
	\label{matrix}
\end{equation}

Figure \ref{cmp_stat} shows the results of the statistics. The panels $A$, $B$ and $C$ are obtained for data of the northern hemisphere (red-coloured plots) and panels $D$, $E$ and $F$ are that of the southern hemisphere (blue-coloured figures). In panels $A$ (northern hemisphere) and $F$ (southern hemisphere), the matrix $M$ (Eq. \ref{matrix}) is visualised. The horizontal axis is the distance from AL. The vertical axis is the logarithm of parameter $S$; $\log  S < 0$ stands for a high-risk flare or CME occurrence. The colour code is the standard score of sunspot group area. The red or blue shades indicate positive standard score ($T_{i} > \overline{T}$). The white color stands for negative standard score ($T_{i} < \overline{T}$ or no data). Panels $B$ and $E$ are the row averages of matrix $M$. Panels $C$ and $D$ are the column averages of matrix $M$.

In Panel $A$, significant islands are visible between $0 > \log S > -0.5$ at $\Delta\psi < 0.1$. There is one more obviously visible island around $\Delta\psi = 0.2$ at $\log S < -0.4$. However, above $\Delta\psi > 0.2$ there is no remarkable island. Panel $C$ of Figure \ref{cmp_stat} also reveals a peak below $\Delta\psi < 0.1$. The statistics suggests, that the most complex and largest sunspot groups appear near the AL.

Analysis of the data of the southern hemisphere data show similar results. There are easily notable islands below $\Delta\psi < 0.2$ at $\log S = -0.6$. Panels $D$ and $E$ of Figure \ref{cmp_stat} clearly reveal that significant sunspot groups appear only near the AL.

Both statistics suggest that the most complex active regions tend to cluster near the AL The co-dominant AL around $\Delta\psi = 0.5$ does not have a significant activity. This statistical investigation also highlights a non-equivalent AL and co-dominant AL activity.

\subsection{Tilt angle of investigated Active Regions within AL}

 \begin{figure}
	\centering
	\includegraphics[width=75mm]{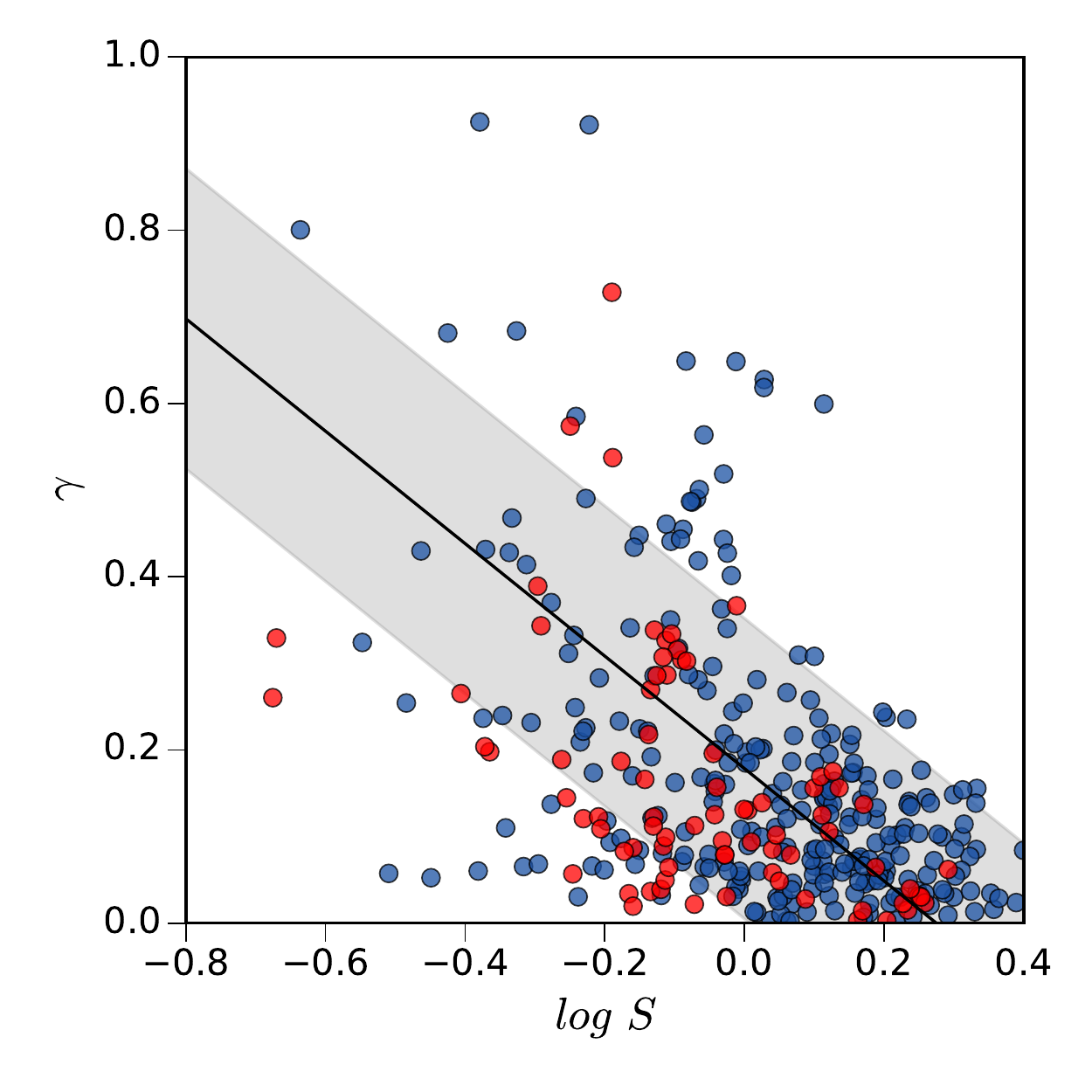}
	\caption{The tilt angle versus separateness parameter of sunspot groups for each hemispheres. Data from northern and souther hemispheres, respectively, are distinguished by blue and red dots. The yellow arrows visualise the result of the PCA method.}
	\label{tilt_stat}
\end{figure}

 \begin{figure}
	\centering
	\includegraphics[width=75mm]{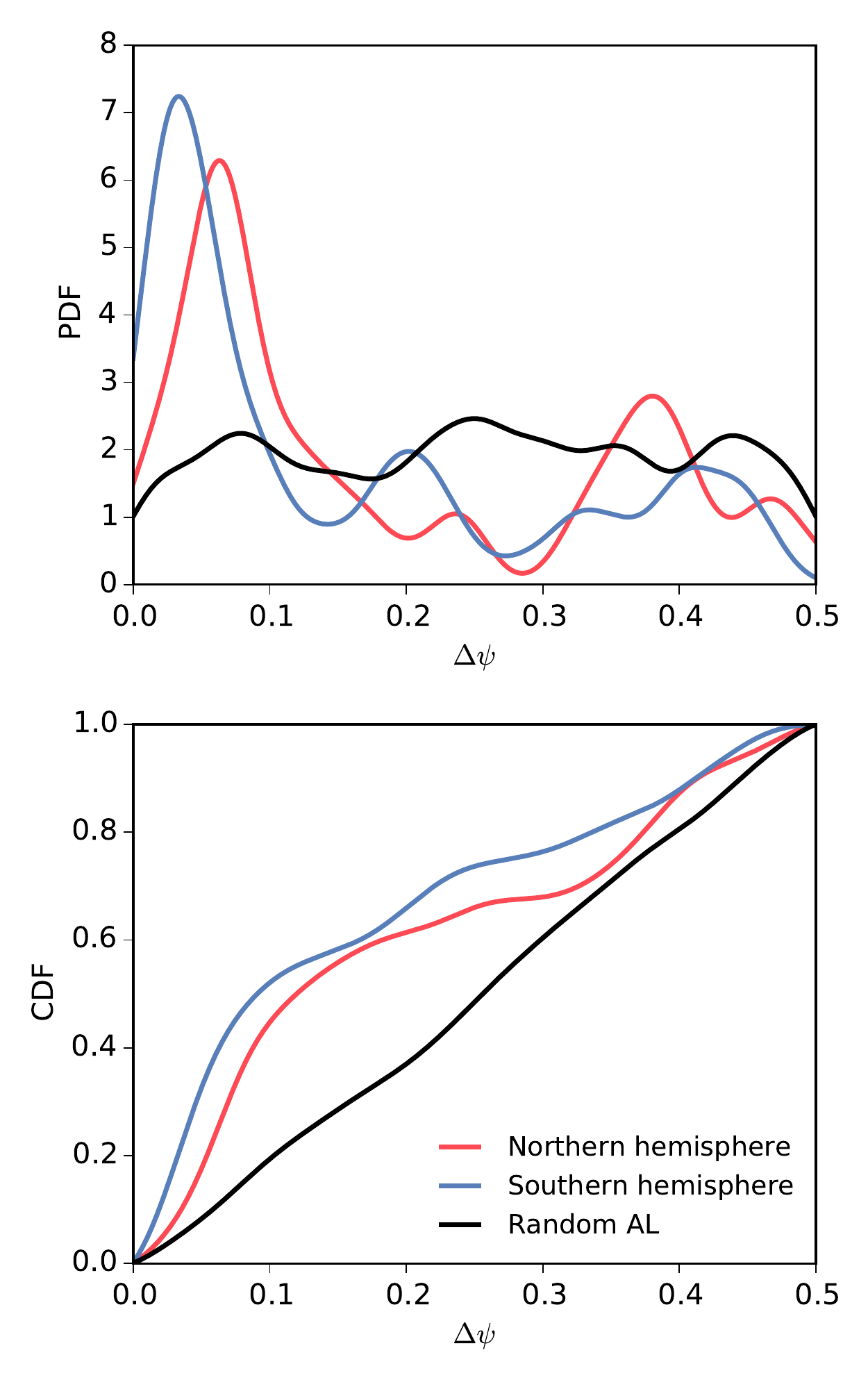}
	\caption{ The upper/lower panel shows the PDF/CDF for CME occurrence. In the upper panel, the blue and red coloured lines are the PDF, obtained by KDE method, distinguished by the northern and southern hemispheres, respectively. The black line is from determining the PDF based on random-generated AL. In the lower panel, the CDFs of the above defined functions are shown.}
	\label{CMEspatial}
\end{figure}

The last investigated morphological property of ARs, here, is the sunspot group tilt angle. The definition of the tilt angle $\gamma^{*}$ is given by \cite{Howard91}; 

\begin{equation}
		\gamma^{*} = \text{atan}\bigg( \frac{(B_{f} - B_{l})/(L_{f} - L_{l})}{\text{sign}(|B_{f} |-|B_{l}|) \cos(B)}  \bigg) .
	\label{tilt}
\end{equation}

Parameters $B$ and $L$ are the Carrington latitude and longitude. The following and leading subgroups have subscripts $f$ and $l$, respectively.

Figure \ref{tilt_stat} demonstrates the relationship found between the separateness parameter  $S$ and scaled tilt angle $\gamma = | \gamma^{*} / 90 |$. Only most significant ARs are taken into account; the area of sunspot groups has to be at least $2\sigma$ greater than average. The results applicable to the two hemispheres are distinguished by the colours of dots.

Linear regression cannot be used because the data is associated with a considerable uncertainty both in the X and Y directions \citep{Isobe90}. For that reason, Principal Component Analysis (PCA) is used to fit the dataset. The PCA method is a linear dimensionality reduction keeping only the most significant singular vectors to project the data to a lower dimensional space \citep{Einbeck07}. The eigenvector of the first component $\boldsymbol{e_{1}} = [-0.8387,  0.5445]$ shows the direction of the maximum variance of the data ($\sigma^2 = 4.1227$), i.e where the data is most spread out.

Based on the result of the PCA we performed Principal Component Regression (PCR). In Figure \ref{tilt_stat}, the solid black line is the regression fit and the grey halo represents a $1\sigma$ standard deviation of the sample along the regression line. The difference between the samples of two hemispheres is statistically insignificant, therefore the regression was applied to the data of both hemispheres. The obtained statistics suggests that there is a clear relationship between the tilt angle and the separateness parameter of the sunspot groups.

\section{Enhanced longitudinal behaviour of CME events}
\subsection{Spatial probably CME occurrences}

In this section, the connection between the CME occurrence and AL is revealed. Panel $A$ of Figure \ref{CMEspatial} shows the kernel probability density function of the longitudinal distribution of CME occurrence. This statistics is based on data from both the northern and southern hemispheres. There is only one significant peak visible around $\Delta\psi = 0.05$. Besides this remarkable peak, there is a long plateau with some insignificant local peaks. Only one more peak, above the significance level at $\Delta\psi = 0.4$, is present, but with a relatively weak activity when compered to the first peak.

A random-generated control group is also used in this statistic. The longitudinal position of AL now is a random position. This test was inspired by \cite{Pelt05} who expressed a critical view on the identification method of \cite{Berdyugina03} employed for AL. In our study, we applied the methodology introduced by \cite{Pelt05}, who reconstructed the distribution of AL with random sunspot longitude data. The KDE plot of the control group does not show any peaks. This homogeneous distribution means that AL identification does not cause false significant peaks, which would affect the results.
 
Panel $B$ of Figure \ref{CMEspatial} shows the cumulative distribution of the above-defined spatial distributions. The blue and red lines have a steep increasing phase between values of $0$ and $0.1$ followed by a less steep increasing trend. These results allow us to estimate that most of CMEs (around $60\%$) occur in a $\pm 36^{\circ}$ belt around the position of AL. Hence, the width of the longitudinal belt of CME occurrences is equal to the width of the longitudinal belt of solar flare occurrences (GY16). The black line is the cumulative distribution obtained from the analysis applied to random longitudinal positions. This distribution would only contain $20\%$ of CMEs. This latter finding means that AL plays a significant role in the spatial distribution of CME occurrences.
 
\subsection{CME dynamics}

  \begin{figure}
	\centering
	\includegraphics[width=85mm]{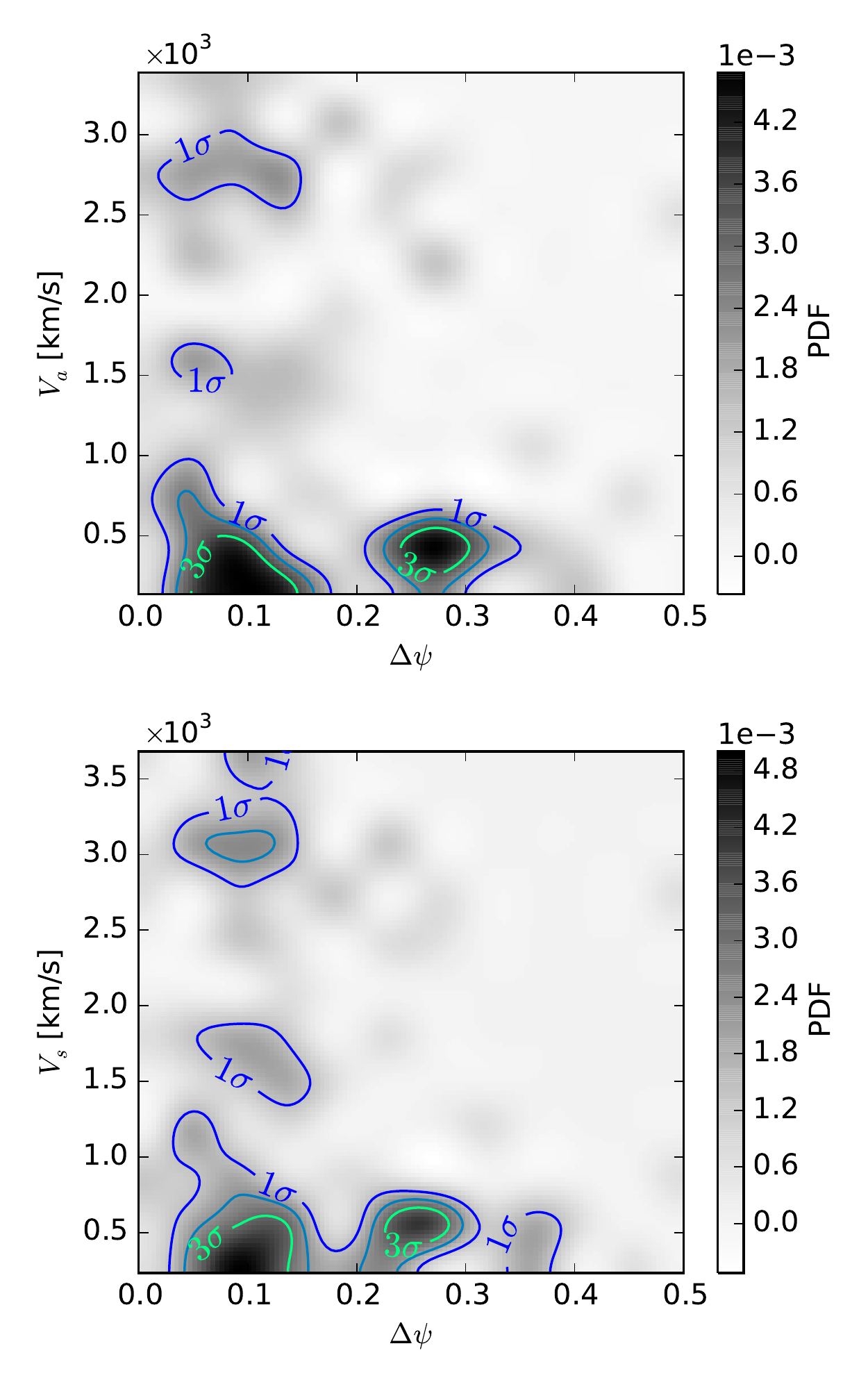}
	\caption{The result of the two-dimensional KDE, using parameters  $\Delta\psi$ and $V_{a}$ (apparent velocity of CME, upper panel),  $V_{s}$ (space velocity of CME, lower panel). The shade of the grey colour represent the probability density. The significant islands are indicated by blue ($1\sigma$), dark green ($2\sigma$) and bright green ($3\sigma$) colours. The northern and souther hemispheres are not distinguished from each other.}
	\label{CMEspatial2}
\end{figure}

Let us now consider the apparent and space velocities ($V_{a}$ and $V_{s}$) of CME events. Two-dimensional kernel density estimations are applied with an axis-aligned bi-variate Gaussian kernel, evaluated on a square grid of the $\Delta\psi - V_{a}$ and $\Delta\psi - V_{s}$ space. Figure \ref{CMEspatial2} shows the result, based on data from both hemispheres. The significance levels $1 - 3\sigma$ are indicated by coloured contour lines.

In both panels of Figure \ref{CMEspatial2}, there are four islands above the $1\sigma$ significance level. The statistics shows that the source of fast CMEs (speeds between $1500$ km/s and $3000$ km/s) is indeed an active region, located within AL. However, slow (i.e. speed less than $1500$ km/s) CMEs can occur outside of AL.

Above the significance level of $3\sigma$, there are only two islands. These are only slow CMEs inside and outside of AL. Analysis of this statistics also indicates that the probability of a slow CME is two standard deviation units higher than the probability that of a fast CME.

\section{Discussion}

The AL identification method presented here reveals new spatial properties of the longitudinal distribution of the sunspot groups (Panels $D$,  $E$,  $F$ of the Fig \ref{al}). The spatial distribution of smaller sunspot groups (less then $1\sigma$) is homogeneous. There is no enhanced longitudinal belt identifiable based on small sunspots; small groups appear everywhere as function of  longitude. Moderate sunspot groups (between $1 - 2\sigma$) show already inhomogeneous properties. However, these results still have to be treated with caution. Only  sunspot groups above the $3\sigma$ significance level have signatures of obvious and remarkable inhomogeneous spatial distribution. 

The idea of two, almost equally significant longitudinal zones is widely accepted by numerous studies, see e.g. \cite{Berdyugina03}. The dominant and co-dominant active longitude is separated by 180 degrees \citep{Zhang11b, Bumba2000}. However, we do not find such equally strong ALs (neither here nor in GY16). In our investigation, the co-dominant AL plays a less important role.

The spatial distribution of the separateness parameter (defined by Eq. \ref{eq5}) shows that complex active regions with a high CME capability appear near mostly the AL (Figure \ref{cmp_stat}). The appearance of moderate and simple complex configurations are everywhere on the solar surface. These groups are also able to have CMEs with a significantly lower probability. 

We also found that, the most tilted sunspot groups have a complex configuration (Figure \ref{tilt_stat}). Simple bipolar sunspot groups show relatively small tilt angle. \cite{Sakurai03} and \cite{Canfield98} concluded, that there is positive correlation between magnetic helicity and sunspot tilt angle. The sunspot rotation could play important role in helicity transport across the photosphere. Sunspot rotation may increase helicity in the corona leading to flares and CMEs \citep{Pevtsov12}. This property may also have the consequence of a more complex built-up of the underlying magnetic structure, and, the well-studied magnetic arches of the upper solar layer could be oriented at a large angle to the equator \citep{Grigorev12}. The more complex active regions are the more flares and they will be associated with CMEs \citep{Huang13, Jetsu97, Kitchatinov05}. Hence, we conclude that the above physical process can take place within AL and anywhere else but only with low probability there.

Several studies have investigated east-west asymmetry of CMEs occurrence. \cite{Skirgiello05} found asymmetry using data provided by the SOHO-LASCO $1996-2004$. The asymmetric behaviour could be a consequence of AL. Our result obtained here shows (see e.g. Fig \ref{CMEspatial2}) that the number of CME occurrence is marginally higher within AL. The mean of the apparent and space velocity of CME occurrences is around $500$ km/s considering the entire surface of the Sun. This mean velocity is known an 'slow' CMEs, found in a number of earlier studies \citep{Ying16}. However, the mean velocity is significantly higher if only the AL itself is considered. Within AL the average (or space) velocity is around $1000$ km/s \citep[see e.g.][]{Michalek09}. There is no fast CME occurrence found outside of AL. Therefore, interestingly and notably, the fast HALO CMEs are also AL CMEs.

\section{Conclusion}

Our new findings (together with the results of GY16) could provide novel aspects both for space weather forecast and for solar dynamo theory. Usually, the flare and/or CME prediction tools are based on only the behaviour of active regions, such as complexity of magnetic fields or other morphological properties. However, the spatial distribution of active regions can also assist in forecasting as suggested by e.g \citet{Zhang08}. We conclude, that the main source of CME and solar flare (GY16) occurrences is the AL. Hence, the detection of this enhanced longitudinal belt may allow us to find the most flare- and CME-capable regions of the Sun preceding the appearance of an active region. This potential flare and CME source is predictable even several solar rotations in advance.

The observed properties of the non-axisymmetric solar activity need to be taken into account in developing and verifying suitable dynamo theory: the observations analysed here show that there is only one significant AL with a relatively wide ($\pm 20-30$ degrees) belt. Furthermore, the tilt angle of the active regions is also an important observed constraint for dynamo theory: the tilt angle of sunspot groups shows non-axisymmetric behaviour, which is a completely new (and surprising) finding.

\section*{Acknowledgments}

The results of this research was enabled partially by SunPy, an open-source and free community-developed Python solar data analysis package \citep{Mumford}. NG thanks for the support received from the University of Sheffield. NG also thanks for the laborious work done by the assistant fellows at Debrecen Heliophysical Observatory for composing the DPD sunspot catalog. RE acknowledges the support received by the Chinese Academy of Sciences President’s International Fellowship Initiative, Grant No. 2016VMA045, the Science and Technology Facility Council (STFC), UK and Royal Society (UK). A.K.S. acknowledges the RESPOND-ISRO (DOS/PAOGIA205-16/130/602) project and the SERB-DST project (YSS/2015/000621) grant.

\end{document}